\documentclass[conference]{IEEEtran}
\newcommand\tab[1][1cm]{\hspace*{#1}}
\usepackage{amsfonts}
\usepackage{xcolor}
\usepackage{adjustbox}
\usepackage{graphicx,booktabs}
\usepackage[cmex10]{amsmath}
\usepackage{dsfont}
\usepackage{cite}
\usepackage{lipsum}
\usepackage{tabu}
\usepackage{amsmath}
\usepackage{bm}
\newcommand{\myMatrix}[1]{\bm{\mathit{#1}}}
\usepackage{xparse}
\usepackage{amsfonts}
\usepackage{amssymb}
\usepackage{resizegather}
\usepackage{algorithmic}
\usepackage[center]{caption}
\usepackage{epsfig}
\usepackage{latexsym}
\usepackage[linesnumbered,ruled,vlined]{algorithm2e}
\usepackage{epstopdf}
\usepackage{amsmath, amsthm, amssymb,enumitem}
\usepackage{verbatim}
\usepackage{amsthm}
\graphicspath{{./figures/}}
\usepackage{graphicx}
\usepackage{subfig}
\usepackage{booktabs}

\usepackage{romannum}
\usepackage{color}
\newtheorem{theorem}{Theorem}

 \author{\IEEEauthorblockN{Ahmed Roushdy\IEEEauthorrefmark{2}\IEEEauthorrefmark{1},
        Abolfazl Seyed Motahari\IEEEauthorrefmark{3}, Mohammed Nafie\IEEEauthorrefmark{1} and Deniz G\"{u}nd\"{u}z\IEEEauthorrefmark{2}
        }
        \IEEEauthorblockA{\IEEEauthorrefmark{2} Information Processing and Communication Lab, Imperial College London, UK\\
    \IEEEauthorblockA{\IEEEauthorrefmark{1}Wireless Intelligent Networks Center (WINC), Nile University, Egypt\\
            \IEEEauthorrefmark{3}EECE Dept., Department of Computer Engineering,
Sharif University of Technology, Iran
    }
} 
        Email: \{\{ahmed.elkordy17, d.gunduz\}@imperial.ac.uk, motahari@sharif.edu, mnafie@ieee.org\}\\
        }
        
\begin{document}
\title{Cache-Aided Fog Radio Access Networks\\
 with Partial Connectivity}
\maketitle
\begin{abstract}
Centralized coded caching  and delivery is studied 
for a partially-connected fog radio access network (F-RAN), whereby a set of $H$  edge nodes
(ENs) (without caches), connected to a cloud server via orthogonal fronthaul links, serve $K$ users over the wireless  edge. The cloud   server is assumed to hold a library of $N$ files, each of size $F$ bits; and each user, equipped with a cache of size $MF$ bits, is connected to a distinct set of $r$  ENs; or equivalently, the wireless edge from the ENs to the users is modeled as a partial interference channel. The objective is  to minimize the normalized delivery time (NDT), which refers to the worst case delivery latency, when each user requests a single file from the library. An achievable coded caching and transmission  scheme is proposed, which  utilizes  maximum distance separable (MDS) codes in the  placement phase, and real interference alignment (IA) in the delivery phase, and its achievable NDT is presented for $r=2$ and arbitrary cache size $M$, and also for arbitrary values of  $r$ when the cache capacity is sufficiently large.
\end{abstract}
\let\thefootnote\relax\footnotetext
{This work was supported in part by the European Union`s Horizon 2020 research and innovation programme under the Marie Sklodowska-Curie action TactileNET (grant agreement No 690893), by the European Research Council (ERC) Starting Grant BEACON (grant agreement No 725731), and by a grant from the Egyptian Telecommunications Regulatory Authority.}

\section{Introduction}
Proactively caching popular contents into user devices  during off-peak traffic periods by exploiting the increasingly abundant storage resources in mobile terminals, has been receiving increasing attention as a promising solution to reduce the increasing  network traffic and latency for 5G and future communication networks. A centralized coded proactive caching scheme is first studied by Maddah-Ali and Niesen in \cite{maddah2014}, where  a single server serves multiple cache-enabled users through an error-free shared link; and it is shown to provide significant coding gains with respect to  classical uncoded caching. More recently, the idea of coded caching has been extended to multi-terminal wireless networks, where transmitters and/or receivers are equipped with cache memories \cite{7282567, 7857805, 2017arXiv170304349P}. It is shown in \cite{7282567} that caches at the transmitters can improve the sum degrees of freedom (DoF) by allowing cooperation between transmitters for interference mitigation. In \cite{7857805} and \cite{2016arXiv160500203X} this model is extended  to a $K_T\times K_R$ network, in which both the transmitters and receivers are equipped with cache memories. An achievable scheme exploiting real interference alignment (IA) for the general $K_T\times K_R$ network is proposed in \cite{2017arXiv170304349P}, which also considers decentralized caching at the users. 

While the aforementioned papers assume that the transmitter caches are large enough to store all the database, the fog-aided radio access network (F-RAN) model, introduced in \cite{sengupta2016cloud}, relaxes this requirement, and allows the delivery of contents from the cloud server to the edge-nodes (ENs) through dedicated fronthaul links. Coded caching for the F-RAN scenario with cache-enabled ENs is studied in \cite{sengupta2016cloud}. The authors propose a centralized  coded caching scheme to minimize  the normalized delivery time (NDT), which measures the worst case delivery latency with respect to an interference-free baseline system in the high signal-to-noise ratio (SNR) regime. In \cite{koh2017cloud}, the authors consider a wireless fronthaul that enables coded multicasting. In \cite{girgis2017decentralized}, decentralized coded caching is studied for an F-RAN architecture with two ENs, in which both the ENs and the users have caches. We note that the models in \cite{sengupta2016cloud, koh2017cloud, girgis2017decentralized} assume a fully connected interference network between the ENs and users. A partially connected F-RAN  is studied in \cite{e19070366} from an online caching perspective.

If each EN is connected  to a subset of the users through dedicated orthogonal links, the corresponding architecture is called a \textit{combination network} \cite{ji2015fundamental, tang2016coded, zewail2017coded}. In combination networks, the server is connected to a set of
$H$ relay nodes (i.e., ENs), which communicate to $K={H \choose r}$ users, such that each user is connected to a distinct set of $r$ relays. The links are assumed to be error and interference free. The objective is to determine the minimal max-link load $R$, defined as the smallest max-rate (the maximum rate among all the links, proportional to the download time) for the worst case demand. Note that, although the delivery from the ENs to the users takes place over orthogonal links,  that is, there are no multicasting  opportunities as in the Maddah Ali and Niesen model in\cite{maddah2014}, the fact that the contents for multiple users are delivered from the server to each relay through a single link, allows coded delivery to offer similar  gains. The authors of \cite{tang2016coded}  consider networks that satisfy the resolvability property, which requires $H$ to be  divisible
by $r$. Combination networks with caches at both the relay and the users is studied  in \cite{zewail2017coded}. For the case when there are no caches at the relays, the authors are able   to achieve the same performance as in \cite{tang2016coded} without requiring the  resolvability property.  A partially connected cache-aided network model is studied in \cite{Mital:WCNC:18}, which assumes a random topology during the delivery phase, which is unknown during placement. 
 
In this paper we study the centralized caching problem in an F-RAN with cache memories at the users as depicted  in Fig. 1. Our work is different from the aforementioned prior works on F-RANs in that, we consider partially connected interference channel from the ENs to the users, instead of a fully connected F-RAN architecture. This may be due to physical constraints that block the signals, or the long distance between some of the EN-user pairs.

Noفe that the considered network  topology, in which the ENs act as relay nodes for the users they serve, is similar to a combination network; however, we consider interfering wireless links  from the ENs to the users instead of dedicated links, and study the normalized delivery time (NDT) in the high SNR regime. The authors in \cite{2017arXiv170809117X} study the NDT for  a partially connected $(K+L-1)\times K$ interference channel with caches at both the transmitters  and the receivers, where each  receiver is  connected   to $L$ consecutive  transmitters. Our work is different from \cite{2017arXiv170809117X}, since we also take into consideration the fronthaul links from the server to the ENs, and consider a network topology in which  the number of transmitters (ENs in our model) is less than or equal to the number of receivers. 

We formulate the minimum NDT problem for an arbitrary \textit{receiver connectivity} $r$, which denotes the number of ENs each user is connected to. Then, we propose a centralized caching and delivery scheme that exploits real interference alignment (IA) to minimize the NDT 
for receiver connectivity of $r=2$. We then extend this scheme to  an arbitrary receiver connectivity $r$ for certain cache capacities. For the proposed scheme, we show that increasing the receiver connectivity, $r$, for the same number of ENs and users will reduce the total NDT for the specific cache capacity region studied, while the amount of reduction depends on the fronthaul capacity. 

\textit{Notation:} We denote sets with calligraphic symbols, and vectors with bold symbols. The set of integers $\{1,\ldots, N\}$ is denoted by $[N]$. The cardinality of set $\mathcal{A}$ is denoted by $|\mathcal{A}|$.

\section{System Model and Performance Measure} \label{system_model} 
 \begin{figure}
\centering
\includegraphics[width=8cm,height=8cm,keepaspectratio]{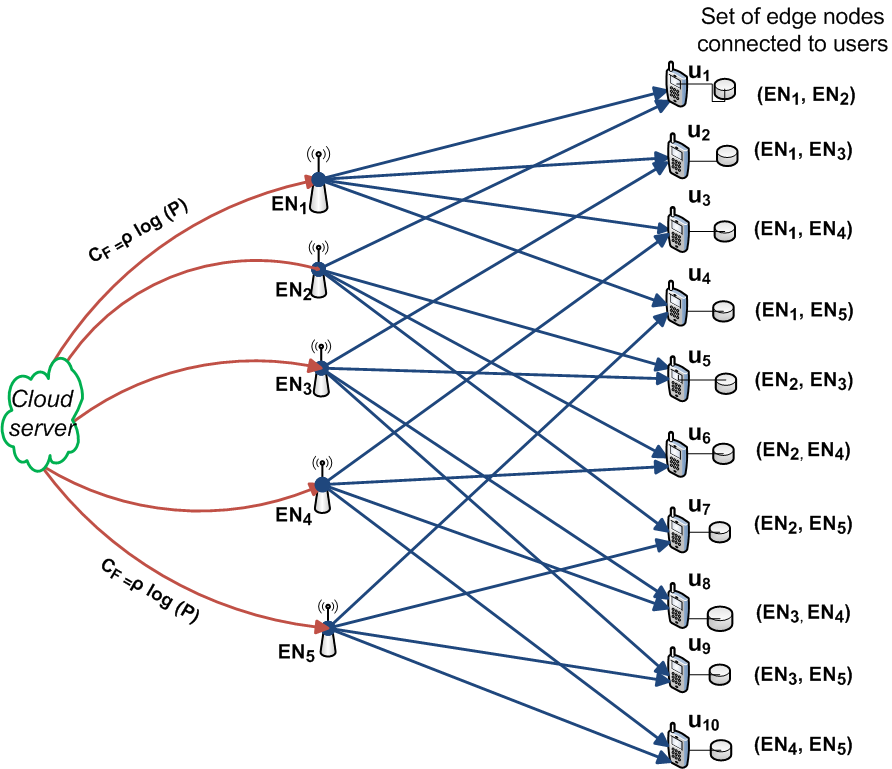}
\caption{F-RAN architecture with receiver connectivity $r=2$, where $H=5$ ENs serve $K = 10$ users.}
  \vspace{-1.5em}
\end{figure}
\subsection{System Model}
We consider an $H\times K$ F-RAN, illustrated in  Fig.~1, which consists of a cloud server that holds a library of $N$ files, $\mathcal{W}\stackrel{\Delta}{=}\{W_{1},\ldots,W_{N}$\}, each of size $F$ bits; a set of $H$ ENs, $\mathcal{R}\stackrel{\Delta}{=}\{\mathrm{EN}_1,\ldots,\mathrm{EN}_H\}$, that help the cloud server  to serve the requests from a set of $K$  users, $\mathcal{U}\stackrel{\Delta}{=}\{\mathrm{u}_1,\ldots,\mathrm{u}_K\}$. The edge network from the ENs to the users is a partially connected interference channel, where each user $\mathrm{u}_k\in\mathcal{U}$ is connected to a distinct set of $r$ ENs, where $r < H$ 
is referred to as the $\textit{receiver connectivity}$. The number of users is $K={H \choose r}$, which means that  $H\leq K$. In  this F-RAN architecture, $\mathrm{EN}_i$, $i\in[H]$,  is connected to $L={H-1\choose r-1}=\frac{rK}{H}$  users.  Each  user is equipped with a cache memory  of size $MF$ bits, while the ENs have no caches. We define the normalized cache capacity of users as $t \triangleq \frac{ML}{N}$.
\setlength{\abovedisplayskip}{4pt}
\setlength{\belowdisplayskip}{3pt}
We denote the set of  users connected to $\mathrm{EN}_i$ by $\mathcal{K}_i$, where $|\mathcal{K}_i|=L$, and the set of  ENs connected to user $\mathrm{u}_k$ by $\mathcal{N}_k$, where  $|\mathcal{N}_k|=r$. We will use the function $\mathrm {Index}(i, k)$ : $[H]\times [K]$ $\,\to\, [L] \cup \{ \epsilon\}$, defined in \cite{zewail2017coded}, which returns $\epsilon $ if $\mathrm{u}_k$ is not served by $\mathrm{EN}_i$, and otherwise returns   the relative order of user $\mathrm{u}_k$  among the users served by $\mathrm{EN}_i$. For example, in Fig. 1, we have $\mathcal{K}_1=\{1, 2, 3, 4\}$, $\mathcal{K}_3=\{2, 5, 8, 9\}$ and
\begin{align*}
\mathrm {Index}(1, 2)&=2, \; \; \; \; \; \mathrm {Index}(1,3)=3, \; \; \; \; \mathrm{Index}(1,5)=\epsilon,\\
\mathrm {Index}(3, 2)&=1, \; \; \; \; \;  \mathrm{Index}(3,5)=2,  \; \; \; \; \mathrm{Index}(3,1)=\epsilon.
\end{align*}

The system operates in two phases: a \textit{placement phase} and a \textit{delivery phase}. The placement phase takes place when the traffic load is low, and  the users are given access to the entire library $\mathcal{W}$. Each user $\mathrm{u}_k$ is then able to fill
 its cache, denoted by $Z_k$, using the library without any prior knowledge of the future 
demands or the  channel coefficients. In the delivery phase, $\mathrm{u}_k$ requests a file $W_{d_k}$ from the library. We define $\mathbf{d}=[d_1,. . . , d_K ] \in [N]^{K}$ as the demand vector. The cloud is connected to each EN via a fronthaul link of capacity $C_F$ bits per symbol, where the symbol refers to a single use of the edge channel from the ENs to the users. 

Once the demands are received, the cloud server sends message $\mathbf{G}_i=(G_i(n))_{n=1}^{T_F}$ of blocklength $T_F$ to $\mathrm{EN}_i$, $i\in[H]$, via the
fronthaul link. This message is limited to $T_FC_F$ bits to guarantee correct decoding at $\mathrm{EN}_i$ with high probability.
 In this paper, we consider half-duplex ENs in that is, ENs start transmitting only after receiving their messages from the cloud server. This is called \textit{serial transmission} in  \cite{sengupta2016cloud}, and the overall latency is the sum of the  latencies in the fronthaul and the edge connections.  $\mathrm{EN}_i$ has an encoding function  that maps the fronthaul message $\mathbf{G}_i$, the demand
vector $\mathbf{d}$, and the channel coefficients $ \mathbf{H} \overset{\Delta}{=}\{h_{k,i}\}_{k\in[K], i\in [H]}$, where $h_{k,i}$ denotes the complex channel gain from $\mathrm{EN}_i$ to $\mathrm{u}_k$,  to a message $\mathbf{V}_i=(V_i(n))_{n=1}^{T_E}$ of blocklength $T_E$, which must satisfy  a power constraint of P.  User $\mathrm{u}_k$ decodes its requested file as  $\hat{W}_{d_k}$ by using its  cache contents $Z_k$, the received message $\mathbf{Y}_k=(Y_k(n))_{n=1}^{T_E}$, as well as its knowledge of the channel gain $\mathbf{H}$ and  the demand vector $\mathbf{d}$.  We have
\begin{equation}
Y_k(n)=\sum_{i \in \mathcal{N}_k} h_{k,i} V_i(n)+n_k(n), 
\end{equation}
where $n_k(n) \sim \mathcal{C N}(0, 1)$ denotes the complex Gaussian noise
at the $k$th user.  The channel gains are independent and identically distributed (i.i.d.) according to a continuous  distribution, and remain constant within each transmission interval. The probability of error for  a coding scheme, consisting of the  caching, cloud
encoding, EN encoding, and user decoding functions,  is defined as 
\begin{equation}
P_e=\max_{\mathbf{d} \in [N]^{K}} \max_{k \in [K]} \mathbb{P}_e( \hat{W}_{d_k} \neq W_{d_k}),
\end{equation}
which is the worst-case probability of error over all possible
demand vectors  and over all users. We say that a coding scheme is
$\textit{feasible}$, if and only if we have $P_e \,\to\, $ 0 when $F\,\to\,  \infty$, for almost all  realizations of the channel matrix $\mathbf{H}$.

\begin{table*}[]
  \centering
 \resizebox{\linewidth}{!}{
    \begin{tabular}{|l|l|l|l|l|l|l|l|l|l|l|}
        \hline
        User & ~ $\mathrm{u}_1$                                    & ~   $\mathrm{u}_2$                                  & ~                                       $\mathrm{u}_3$ & ~                 $\mathrm{u}_4$                   & ~         $\mathrm{u}_5$                             & ~     $\mathrm{u}_6$ & ~ $\mathrm{u}_7$ & ~    $\mathrm{u}_8$                                  & ~  $\mathrm{u}_9$                                   & ~                                             $\mathrm{u}_{10}$ \\ \hline
        Cache Contents   & $f_{n,1}^1$, $f_{n,1}^2$  & $f_{n,2}^1$, $f_{n,1}^3$   & $f_{n,3}^1$, $f_{n,1}^4$     & $f_{n,4}^1$, $f_{n,1}^5$   & $f_{n,2}^2$, $f_{n,2}^3$    & $f_{n,3}^2$, $f_{n,2}^4$   & $f_{n,4}^2$, $f_{n,2}^5$  & $f_{n,3}^3$, $f_{n,3}^4$    & $f_{n,4}^3$, $f_{n,3}^5$   & $f_{n,4}^4$, $f_{n,4}^5$           \\
        \hline
    \end{tabular}
    }
  \caption{Cache contents after the placement phase for the F-RAN scenario considered in Example 1, where $K=N=10$, $r=2$, $L=4$,  $t=1$ and $M=\frac{5}{2}$. }
  \label{Table:a}
  
\end{table*}

\begin{table*}[]
\centering
\begin{tabular}{|l|l|l|l|l|}
\hline
\multicolumn{1}{|c|}{$\mathrm{EN}_1$}                                     & \multicolumn{1}{c|}{$\mathrm{EN}_2$}                 & \multicolumn{1}{c|}{$\mathrm{EN}_3$}                & \multicolumn{1}{c|}{$\mathrm{EN}_4$}                 & \multicolumn{1}{c|}{$\mathrm{EN}_5$}                 \\ \hline
\multicolumn{1}{|c|}{$\mathbf{X}_{1}^{1,2}=f_{1,2}^1+f_{2,1}^1$} & $\mathbf{X}_{2}^{1,2}=f_{1,2}^2+f_{5,1}^2$ & $\mathbf{X}_{3}^{1,2}=f_{2,2}^3+f_{5,1}^3$ & $\mathbf{X}_{4}^{1,2}=f_{3,2}^4+f_{6,1}^4$  & $\mathbf{X}_{5}^{1,2}=f_{4,2}^5+f_{7,1}^5$  \\
$\mathbf{X}_{1}^{1,3}=f_{1,3}^1+f_{3,1}^1$                       & $\mathbf{X}_{2}^{1,3}=f_{1,3}^2+f_{6,1}^2$  & $\mathbf{X}_{3}^{1,3}=f_{2,3}^3+f_{8,1}^3$ & $\mathbf{X}_{4}^{1,3}=f_{3,3}^4+f_{8,1}^4$  & $\mathbf{X}_{5}^{1,3}=f_{4,3}^5+f_{9,1}^5$  \\
$\mathbf{X}_{1}^{1,4}=f_{1,4}^1+f_{4,1}^1$                       & $\mathbf{X}_{2}^{1,4}=f_{1,4}^2+f_{7,1}^2$  & $\mathbf{X}_{3}^{1,4}=f_{2,4}^3+f_{9,1}^3$ & $\mathbf{X}_{4}^{1,4}=f_{3,4}^4+f_{10,1}^4$ & $\mathbf{X}_{5}^{1,4}=f_{4,4}^5+f_{10,1}^5$ \\
$\mathbf{X}_{1}^{2,3}=f_{2,3}^1+f_{3,2}^1$                       & $\mathbf{X}_{2}^{2,3}=f_{5,3}^2+f_{6,2}^2$  & $\mathbf{X}_{3}^{2,3}=f_{5,3}^3+f_{8,2}^3$ & $\mathbf{X}_{4}^{2,3}=f_{6,3}^1+f_{8,2}^4$  & $\mathbf{X}_{5}^{2,3}=f_{7,3}^5+f_{9,2}^5$  \\
$\mathbf{X}_{1}^{2,4}=f_{2,4}^1+f_{4,2}^1$                       & $\mathbf{X}_{2}^{2,4}=f_{5,4}^2+f_{7,2}^2$  & $\mathbf{X}_{3}^{2,4}=f_{5,4}^3+f_{9,2}^3$ & $\mathbf{X}_{4}^{2,4}=f_{6,4}^1+f_{10,2}^4$ & $\mathbf{X}_{5}^{2,4}=f_{7,4}^5+f_{10,2}^5$ \\
$\mathbf{X}_{1}^{3,4}=f_{3,4}^1+f_{4,3}^1$                       & $\mathbf{X}_{2}^{3,4}=f_{6,4}^2+f_{7,3}^2$  & $\mathbf{X}_{3}^{3,4}=f_{8,4}^3+f_{9,3}^3$ & $\mathbf{X}_{4}^{3,4}=f_{8,4}^1+f_{10,3}^4$ & $\mathbf{X}_{5}^{3,4}=f_{9,4}^5+f_{10,3}^5$ \\ \hline
\end{tabular}
\caption{The  data delivered  from the cloud server  to each EN for Example 1.}
\label{Table:b}
\end{table*}
\setlength{\textfloatsep}{1cm}
\subsection{Performance Measure}

We will consider the normalized delivery time (NDT) in the high SNR regime \cite{7864374}  as the  performance measure. For cache capacity $M$ and fronthaul capacity $C_F = \rho \log P$, $\delta (N,M, \rho)$ is an \textit{achievable} NDT if there exists a sequence of feasible codes that satisfy 
\begin{equation}
\delta (N,M, \rho) = \lim_{P, F\to \infty}\sup{ \frac{(T_F+T_E)\log P } {F}}.
\end{equation}
We define the minimum NDT for a
given tuple $(N,M, \rho)$ as
\begin{equation*}
\delta^\star (N,M, \rho)=\inf \{ \delta (N,M, \rho) : \delta (N,M, \rho) \text{ is achievable}\}.
\end{equation*}
Let $R_1$  denote the worst-case traffic load from the cloud server to the $\mathrm{EN}_i$, while  $R_2$ denote the worst-case traffic load per
user, both normalized by file size $F$. The per-user capacity in the high SNR regime can be approximated by  $d  \log P + o(\log P)$, where $d$ is the per-user   DoF, while the capacity of the  fronthaul link is given by $\rho \log P $+ o($\log P$), where $\rho\geq0$ is called the \textit{fronthaul rate}. Then,
NDT can be expressed more  conveniently as \cite{2016arXiv160500203X}
\begin{equation}
\delta (N,M, \rho)=\frac{R_1}{\rho}+\frac{R_2}{d}, 
\end{equation}
where $\delta_F\triangleq \frac{R_1}{\rho}$ represents the fronthaul NDT,  and $\delta_E\triangleq \frac{R_2}{d}$ represents the  edge NDT,  which suggests that NDT characterizes the delivery time of the actual traffic load at a transmission rate specified by DoF $d$.

\section{Main Result} \label{MR}
 The main result of the paper is presented next.
 \begin{theorem} For an  $H\times K$ partially-connected F-RAN architecture with user cache capacity of $M$, fronthaul rate  $\rho\geq 0$, number of files $N\geq K$ and centralized cache placement, the following NDT is achievable for integer values of $t\triangleq \frac{ML}{N}$:
\begin{equation}
\label{ax}
 \delta(N,M, \rho) = \frac{L-t}{r}\left( \frac{r-1}{L} +  \frac{1}{t+1}\left(1+\frac{1}{\rho} \right) \right)
 \end{equation}
for a receiver connectivity of $r=2$, or for arbitrary receiver connectivity when $t\geq L-2$. The NDT for non-integer $t$ values can be obtained as a linear combination of the NDTs of integer $t$ values through memory-sharing.
\end{theorem}


\textbf{Remark} From Theorem 1,  when $r>1$, we have
\begin{equation*}
\label{bb}
\delta (N,M, \rho)=\begin{cases}\frac{2}{ r} \left(\frac{r-1}{L}+ \frac{1}{L-1} \left(1+\frac{1}{\rho} \right) \right), & t=L-2\\\ \frac{1}{L} \left( 1+ \frac{1}{\rho r}\right), &  t=L-1\end{cases}. 
\end{equation*}
 Consider two different F-RAN architectures with $H$ ENs, F-RAN A and F-RAN B, with receiver connectivities $r_A$ and $r_B$, respectively, where $r_A + r_B = H$ and $r_A\geq r_B$. The two networks have the same number of users, and we have $L_x=\frac{K}{H}r_x$, $x\in\{A,B\}$. One can then show that the achievable NDT in F-RAN A is lower, showing that the increased connectivity  helps in reducing the NDT despite increasing interference, and  the gap between the two achievable NDTs  in F-RAN A and F-RAN B becomes negligable as the fronthaul rate increases, i.e., $\rho\rightarrow \infty$.   We illustrate the achievable NDT performance in a $7\times 21$ F-RAN    in Fig. 2   for $r_A=5$, $r_B=2$, for  different fronthaul rates. We observe from the figure that,   with the same cache  capacity $M$  the achievable NDT of network A is less than or equal to  that of network B, and the gap between the  two increases as the fronthaul rate decreases.   This suggests that the achievable NDT for a given F-RAN architecture decreases as the connectivity increases where the amount of decreasing depends on the fronthaul rate.

\begin{figure}
\centering
\includegraphics[width=8cm,height=8cm,keepaspectratio]{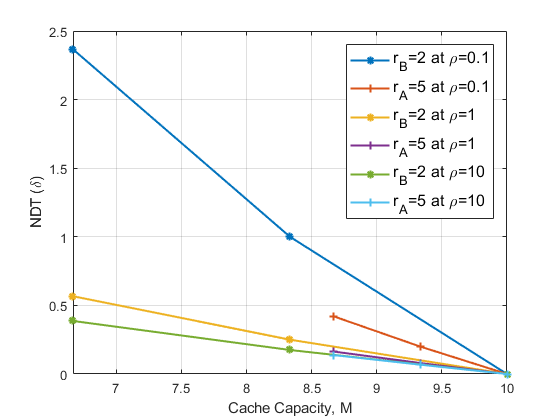}
\caption{Comparison of the  NDT for a $7\times21$ F-RAN architecture  with a library of $N=21$ files, considering different receiver connectivities ($r$) and fronthaul rates ($\rho$).}
\vspace{-1.5em}
\label{xx}
\end{figure}

\section{Centralized Coded Caching}\label{Scheme}

In this section, we  present a centralized coded caching scheme for the partially-connected F-RAN architecture with a receiver connectivity of $r=2$ for $t \in [L]$, and also for  any receiver connectivity $r$ for  $t\geq L-2$. 

\subsection{Cache Placement Phase}
We use the cache placement algorithm proposed in \cite{zewail2017coded}, where the cloud server divides each file $W_n\in \mathcal{W}$ into  $r$ equal size subfiles.  Then, it encodes each subfile using an $(H, r)$ maximum distance separable (MDS) code \cite{Lin:2004:ECC:983680}. The resulting coded chunks, each of size $F/r$ bits,  are denoted by $f_n^i$, where $n$ is the file index and $i\in[H]$. 
Each $\mathrm{EN}_i$ acts as a virtual server for the resulting encoded symbol $f_n^i$. Note that, any $r$ encoded chunks are sufficient to reconstruct the whole file.
 
Each encoded symbol $f_n^i$ is  further divided into ${L\choose t}$ equal-size non-overlapping pieces, each of  which is denoted by $f^i_{n, \mathcal{T}}$, where  $\mathcal{T}\subseteq{\{1, \ldots, L\}}$, $|\mathcal{T}|=t$. The pieces  $f^i_{n, \mathcal{T}}$, $\forall n$, are stored in the cache memory of user $\mathrm{u}_k$ if $k\in  \mathcal{K}_i$ and $\mathrm{Index}(i, k)\in \mathcal{T}$; that is, the pieces of chunk $i$, $i\in[H]$, are stored by users connected to $\mathrm{EN}_i$. At the end of the  placement phase, each user stores
$Nr\binom {L-1} {t-1}$ pieces, each of size $\frac{F}{r{L\choose t}}$
bits, which sum up to $MF$ bits,  satisfying the memory constraint. We will explain the placement phase through an  example.

\textbf{Example 1.}
 $\;$ Consider the  $5\times 10$ partially connected F-RAN depicted in Fig. 1, where $H=5$, $N=10$,  $r=2$ and $L=4$. For $t=1$, i.e., $M=N/L$, the cloud server first  divides each file  into $r=2$ subfiles. These subfiles are then  encoded  using a $(5, 2)$ MDS code. As a result, there are $5$ encoded chunks, denoted by  $f_n^i$, $n\in[10]$, $i\in[5]$, each of size $F/2$ bits. Each encoded chunk  $f_n^i$ is further divided into $4$ pieces $f^i_{n, \mathcal{T}}$, where $\mathcal{T}\subseteq{\{1, \ldots, 4\}}$ and $|\mathcal{T}|=1$. Cache contents of each user are listed in TABLE \ref{Table:a}. Observe that each user stores two 
pieces of the encoded chunks of each file for a total of 10 files, i.e., $\frac{5}{2}F$ bits, which
satisfies the memory constraint.
\subsection{Delivery Phase}
The delivery phase is carried out in two steps. The first step  is the delivery from the  cloud server to the ENs, and the second step is the delivery from the ENs to the users.

\subsubsection{Delivery from the  cloud server to the  ENs}
For each  $(t+1)$-element subset $\mathcal{S}$ of $[L]$, i.e., $\mathcal{S}\subseteq[L]$ and $|\mathcal{S}|=t+1$, the cloud server will deliver the following message to $\mathrm{EN}_i$:
\begin{equation}
\textbf{X}_i^\mathcal{S} \triangleq \bigoplus_{k: k\in \mathcal{K}_i, Index(i, k)\in \mathcal{S}} f^i_{{d_k}, \mathcal{S} \backslash \mathrm{Index}(i, k)}.
\end{equation}
Overall, for  given  $\mathbf{d}$, the following set of messages will be delivered  to $EN_i$
\begin{equation}
\{\textbf{X}_i^\mathcal{S}: \mathcal{S}\subseteq[L], |\mathcal{S}|=t+1\},
\end{equation}
which is of size  $ {L\choose t+1} \frac{F}{r{L\choose t}}$ bits.
The message to be delivered to each  EN in Example 1 is given in TABLE \ref{Table:b}, which results in a normalized fronthaul  traffic load of $R_1=\frac{ {4\choose 2}}{2{4\choose 1}} =\frac{3}{4}$. Hence, the achievable  NDT from the cloud server to the ENs is $\delta_F= \frac{3}{4\rho}$.  In  general  the NDT from the cloud server to the ENs is 
\begin{equation}
 \delta_{F}=\frac{{L\choose t+1} }{r{L\choose t} \rho} = \frac{L-t}{(t+1)r\rho}.
\end{equation}
\begin{table*}[]
\centering
\begin{tabular}{|l|l|l|l|l|l|l|l|l|l|}
\hline
\multicolumn{1}{|c|}{$\mathds{X}_1$}                                          & \multicolumn{1}{c|}{$\mathds{X}_2$}                     & \multicolumn{1}{c|}{$\mathds{X}_3$}                     & \multicolumn{1}{c|}{$\mathds{X}_4$}                     & \multicolumn{1}{c|}{$\mathds{X}_5$}                     & \multicolumn{1}{c|}{$\mathds{X}_6$}                     & \multicolumn{1}{c|}{$\mathds{X}_7$}                     & \multicolumn{1}{c|}{$\mathds{X}_8$}                     & \multicolumn{1}{c|}{$\mathds{X}_9$}                     & \multicolumn{1}{c|}{$\mathds{X}_{10}$}                    \\ \hline
\multicolumn{1}{|c|}{$\mathbf{X}_{1}^{2,3} \; \mathbf{X}_{2}^{2,3}$} & $\mathbf{X}_{1}^{1,3} \; \mathbf{X}_{3}^{2,3}$ & $\mathbf{X}_{1}^{1,2} \; \mathbf{X}_{4}^{2,3}$ & $\mathbf{X}_{1}^{1,2} \; \mathbf{X}_{5}^{2,3}$ & $\mathbf{X}_{2}^{1,3} \; \mathbf{X}_{3}^{1,3}$ & $\mathbf{X}_{2}^{1,2} \; \mathbf{X}_{4}^{1,3}$ & $\mathbf{X}_{2}^{1,2} \; \mathbf{X}_{5}^{1,3}$ & $\mathbf{X}_{3}^{1,2} \; \mathbf{X}_{4}^{1,2}$ & $\mathbf{X}_{3}^{1,2} \; \mathbf{X}_{5}^{1,4}$ & $\mathbf{X}_{4}^{1,2} \; \mathbf{X}_{5}^{1,2}$ \\
$\mathbf{X}_{1}^{2,4} \; \mathbf{X}_{2}^{2,4}$                       & $\mathbf{X}_{1}^{1,4} \; \mathbf{X}_{3}^{2,4}$ & $\mathbf{X}_{1}^{1,4} \; \mathbf{X}_{4}^{2,4}$ & $\mathbf{X}_{1}^{1,3} \; \mathbf{X}_{5}^{2,4}$ & $\mathbf{X}_{2}^{1,4} \; \mathbf{X}_{3}^{1,4}$ & $\mathbf{X}_{2}^{1,4} \; \mathbf{X}_{4}^{1,4}$ & $\mathbf{X}_{2}^{1,3} \; \mathbf{X}_{5}^{1,4}$ & $\mathbf{X}_{3}^{1,4} \; \mathbf{X}_{4}^{1,4}$ & $\mathbf{X}_{3}^{1,3} \; \mathbf{X}_{5}^{1,4}$ & $\mathbf{X}_{4}^{1,3} \; \mathbf{X}_{5}^{1,3}$ \\
$\mathbf{X}_{1}^{3,4} \; \mathbf{X}_{2}^{3,4}$                       & $\mathbf{X}_{1}^{2,4} \; \mathbf{X}_{3}^{3,4}$ & $\mathbf{X}_{1}^{3,4} \; \mathbf{X}_{4}^{3,4}$ & $\mathbf{X}_{1}^{2,3} \; \mathbf{X}_{5}^{3,4}$ & $\mathbf{X}_{2}^{3,4} \; \mathbf{X}_{3}^{3,4}$ & $\mathbf{X}_{2}^{2,4} \; \mathbf{X}_{4}^{3,4}$ & $\mathbf{X}_{2}^{2,3} \; \mathbf{X}_{5}^{3,4}$ & $\mathbf{X}_{3}^{2,4} \; \mathbf{X}_{4}^{2,4}$ & $\mathbf{X}_{3}^{2,3} \; \mathbf{X}_{5}^{2,4}$ & $\mathbf{X}_{4}^{2,3} \; \mathbf{X}_{5}^{2,3}$ \\ \hline
\end{tabular}
\caption{The interference matrices at the  users of  Example 1.}
\label{Table:3}
\end{table*}
\begin{algorithm}[t]
\caption{  Generator for $\mathds{A}$, $\mathds{B}$ and $\mathds{C}$ Matrices}
 $\mathds{A}=[\;]$,   $\mathds{B}=[\;]$,  $\mathds{C}=[\;]$, $g=0$ \\
 FOR  $k=1, . ., K$\\
\tab[.25cm] FOR  $j=1, . ., I$\\
\tab[.375cm]$g=g+1$\\
\tab[.5cm] FOR  $i=1, . ., r$\\
\tab[.75cm] $\mathds{B}_g \leftarrow [\mathds{X}_k(j,i) \; \; \mathds{B}_g]$   \\
\tab[.75cm] Find $\mathcal{J}_i$: set of other users receiving the same 
\tab[.65cm] interference signal $\mathds{X}_k(j,i)$, $|\mathcal{J}_i|=(L-|S|-1)$. \tab[.65cm] Sort users in $\mathcal{J}_i$ in ascending order.\\
 \tab[.75cm]For each user in $\mathcal{J}_i$, find interference vector $\mathbf{x}_k^q$,  \tab[.60cm]s.t. 
 $\mathrm{u}_k\in \mathcal{J}_i$ and $\mathds{X}_k(j,i)\not\in \mathbf{x}_k^q$. \\  \tab[.75cm] $\mathds{Q}_i \leftarrow$ set of vectors $\mathbf{x}_k^q$ \\
\tab[.5cm] END FOR\\
 \tab[1.10cm]If $|\mathcal{J}_i|\geq 1$\\
\tab[1.40cm] FOR $R=1,\ldots , |\mathcal{J}_i|$\\
 \tab[1.70cm]  FOR $e=1, \ldots, |\mathds{Q}_1(:,R)|$  $\; \;$ \\
\tab[2.20cm]FOR $c=1, \ldots, |\mathds{Q}_2(:,R)|$\\
 \tab[2.70cm]IF  $\mathds{Q}_1(e,R) =\mathds{Q}_2(c,R)$\\
 \tab[3cm]$\mathds{B}_g=[   \mathds{B}_g\; \; \mathds{Q}_1(e,R)]$\\
\tab[3cm]Go to 21, i.e., next \tab[2.75cm] iteration of $R$.\\
 \tab[2.70cm]END IF\\
 \tab[2.20cm] END FOR\\
   \tab[1.70cm] END FOR\\
 \tab[1.40cm] END FOR\\
  \tab[1.10cm] END IF\\
\tab[.25cm] $\mathds{C}_g \leftarrow \bigcup\limits_{k:\mathcal{\hat{S}}\in \mathds{X}_k} \mathrm{u}_k,$ for  $\hat{S}\subseteq \mathds{B}_g$, where $|\hat{S}| = r$\\
\tab[3.10cm]   FOR $e=1, \ldots, |\mathds{C}_g|$\\
  \tab[3.40cm] FOR  $i=1, \ldots, r$\\
\tab[3.40cm]$\mathds{A}_g=[\mathds{A}_g \; \;h_{\mathds{C}_g(e),\mathcal{N}_{\mathds{C}_g(e)}(i)}]$\\
\tab[3.40cm]END FOR\\
 \tab[3.10cm]END FOR\\
  \tab[.25cm] Remove interference signals in $\mathds{B}_g$ from $ (\mathds{X}_k)_{k=1}^K$\\
 \tab[.25cm]$\mathcal{J}_i=[\;] \; \; \mathds{Q}_i=[\;]$ $\; \;$ for $\;i= 1, \ldots, r$\\
 \tab[.25cm]END FOR\\
 END FOR\\
\end{algorithm}
\setlength\floatsep{.25\baselineskip plus 3pt minus 2pt}
\setlength\textfloatsep{.25\baselineskip plus 3pt minus 2pt}
\setlength\intextsep{.25\baselineskip plus 3pt minus 2pt}
\subsubsection{Delivery from the ENs to the  users}
User $\mathrm{u}_k$, $k\in[K]$, is interested in the following set of  messages:
\begin{equation}
\mathcal{M}_k=\bigcup\limits_{\substack{i, \mathcal{S}: i \in \mathcal{N}_k, \mathcal{S}\subseteq\{1, \ldots, L\}, \\  \;|\mathcal{S}|=t+1, \; {\mathrm{Index}(i,k)\in \mathcal{S}} }} X_{i}^\mathcal{S},
\end{equation}
where $|\mathcal{M}_k |=r{L-1\choose t}$. On the other hand, the transmission of the following messages interfere with the transmissions of the messages in $\mathcal{M}_k$:
\begin{equation}
\mathcal{I}_k=\bigcup\limits_{\substack{i,  \mathcal{S}: i \in \mathcal{N}_k,  \mathcal{S}\subseteq\{1, \ldots, L\}, \\ \;|\mathcal{S}|=t+1, \; {\mathrm{Index}(i,k)\not \in  S} }} X_{i}^S.
\end{equation}

Each  $X_{i}^S\in \mathcal{I}_k$ causes interference at $L-|\mathcal{S}|$ users,  including  $\mathrm{u}_k$.
 Hence, the  total number of interfering signals at   $\mathrm{u}_k$ from the ENs in   $ \mathcal{N}_k$ is  $rI$, where $I \triangleq {L\choose t+1}-{L-1\choose t}$ is the number of interfering signals from each  EN connected to user $\mathrm{u}_k$.

At each  user $\mathrm{u}_k$, $k \in [K]$, we define  the interference matrix $\mathds{X}_k$  to be a matrix with  $r$ columns,  denoted  by  $\{\mathbf{x}_k^q\}_{q=1}^{r}$, each column representing  the interference caused by  $EN_i\in \mathcal{N}_k$,   and $I$ rows. For  each column vector $\mathbf{x}_k^q$, we sort the set of interfering signals $ \mathcal{I}_k$ for $i=\mathcal{N}_k(q)$ in ascending order, where $\mathcal{N}_k(q)$  is  the q-th element of the set $\mathcal{N}_k$ when they are ordered in ascending order. For Example 1, the interference matrices  are  shown  in  TABLE \ref{Table:3}.

We will use real IA,  presented in ~\cite{6846359} and \cite{5513550},  for the delivery phase from the ENs to the  users to align each of the $r$ interfering signals  in $\mathcal{I}_k$, one from each EN, to the same subspace. We define $\mathds{A}$,  $\mathds{B}$  and  $\mathds{C}$ to be the basis matrix, i.e., function of the channel coefficients, the data matrix and user matrix, respectively, where  the dimensions of these matrices are $G\times r{(r+L-|S|-1)\choose r}$, $G\times (r+L-|S|-1)$ and $G\times {(r+L-|S|-1)\choose r}$, respectively, where $G={H \choose t+1}$. We denote  the rows of these matrices  by $\mathds{A}_g$,  $\mathds{B}_g$ and   $\mathds{C}_g$, respectively, where $g\in[G]$. The row vectors  $\{\mathbf{\mathds{A}}_g\}_{g=1}^G$   are used to generate the set of monomials $\mathcal{G}(\mathds{A}_g)_{g=1}^G$. Note that, the function $\mathcal{T}(u)$ defined in  \cite{7282567} corresponds to $\mathcal{G}(\mathds{A}_g)$ in our notation.  The set $\mathcal{G}(\mathds{A}_g)_{g=1}^G$ is used as the transmission directions for the modulation constellation $\mathbb{Z}_Q$ ~\cite{7282567} for the whole network. In other words,  each row data vector $\mathds{B}_g$  will use the set $\mathcal{G}(\mathds{A}_g)$ as the transmission directions of all its data to align all the  $r$ interfering signals from  $\mathds{B}_g$ at the same subspace at  $\mathrm{u}_k\in \mathds{C}_g$, if these $r$ signals   belong to  $\mathds{X}_k$. 

We next explain matrix $\mathds{C}$ more clearly. For each $\mathcal{\hat{S}}\subseteq \mathds{B}_g$ with $|\mathcal{\hat{S}}| = r$, there will be a  user at which these data will be aligned into the same dimension, i.e., $|\mathds{C}_g|={(r+L-|S|-1)\choose r}$. The row $\mathds{C}_g$ is given as follow,
\begin{equation}
\mathds{C}_g=\bigcup\limits_{k:\mathcal{\hat{S}}\in \mathds{X}_k} \mathrm{u}_k.
\end{equation}  
We employ  Algorithm 1 to obtain matrices $A$, $B$ and $C$ for a receiver connectivity of $r=2$, and for arbitrary receiver connectivity when $t= L-2$. For Example 1, the three matrices  are given  as follows:
\[
\myMatrix{\mathds{A}} = \begin{pmatrix}
h_{1,1} &h_{1,2}  &h_{4,1}&h_{4,5} &h_{7,2}&h_{7,5}\\
h_{1,1} &h_{1,2}  &h_{3,1}&h_{3,4} &h_{6,2}&h_{6,4}\\
h_{1,1} &h_{1,2}  &h_{2,1}&h_{2,3} &h_{5,2}&h_{5,3}\\
h_{2,1} &h_{2,3}  &h_{4,1}&h_{4,5} &h_{9,3}&h_{9,5}\\
h_{2,1} &h_{2,3}  &h_{3,1}&h_{3,4} &h_{8,3}&h_{8,4}\\
h_{3,1} &h_{3,4}  &h_{4,1}&h_{4,5} &h_{10,4}&h_{10,5}\\
h_{5,2} &h_{5,3}  &h_{7,2}&h_{7,5} &h_{9,3}&h_{9,5}\\
h_{5,2} &h_{5,3}  &h_{6,2}&h_{6,4} &h_{8,3}&h_{8,4}\\
h_{6,2} &h_{6,4}  &h_{7,2}&h_{7,5} &h_{10,4}&h_{10,4}\\
h_{8,3} &h_{8,4}  &h_{10,4}&h_{10,5} &h_{9,3}&h_{9,5}\\
\end{pmatrix},
\]
\[
\myMatrix{\mathds{B}} = \begin{pmatrix}
\mathbf{X}_{1}^{2,3}  &\mathbf{X}_{2}^{2,3} & \mathbf{X}_{5}^{3,4} \\
\mathbf{X}_{1}^{2,4}  &\mathbf{X}_{2}^{2,4} & \mathbf{X}_{4}^{3,4} \\
\mathbf{X}_{1}^{3,4}  &\mathbf{X}_{2}^{3,4} & \mathbf{X}_{3}^{3,4} \\
\mathbf{X}_{1}^{1,3}  &\mathbf{X}_{3}^{2,3} & \mathbf{X}_{5}^{2,4} \\
\mathbf{X}_{1}^{1,4}  &\mathbf{X}_{3}^{2,4} & \mathbf{X}_{4}^{2,4} \\
\mathbf{X}_{1}^{1,2}  &\mathbf{X}_{4}^{2,3} & \mathbf{X}_{5}^{2,3} \\
\mathbf{X}_{2}^{1,3}  &\mathbf{X}_{3}^{1,3} & \mathbf{X}_{5}^{1,4} \\
\mathbf{X}_{2}^{1,4}  &\mathbf{X}_{3}^{1,4} & \mathbf{X}_{4}^{1,4} \\
\mathbf{X}_{2}^{1,2}  &\mathbf{X}_{4}^{1,3} & \mathbf{X}_{5}^{1,3} \\
\mathbf{X}_{4}^{1,2}  &\mathbf{X}_{3}^{1,2} & \mathbf{X}_{5}^{1,2} \\
\end{pmatrix}
, \;
\myMatrix{\mathds{C}} = \begin{pmatrix}
\mathrm{u}_1  &\mathrm{u}_4 & \mathrm{u}_7 \\
\mathrm{u}_1  &\mathrm{u}_3 & \mathrm{u}_6  \\
\mathrm{u}_1  &\mathrm{u}_2 & \mathrm{u}_5  \\
\mathrm{u}_2  &\mathrm{u}_4 & \mathrm{u}_9  \\
\mathrm{u}_2  &\mathrm{u}_3 & \mathrm{u}_8  \\
\mathrm{u}_3  &\mathrm{u}_4 & \mathrm{u}_{10}  \\
\mathrm{u}_5  &\mathrm{u}_7 & \mathrm{u}_9  \\
\mathrm{u}_5  &\mathrm{u}_6 & \mathrm{u}_8  \\
\mathrm{u}_6  &\mathrm{u}_7 & \mathrm{u}_{10}  \\
\mathrm{u}_8  &\mathrm{u}_{10} & \mathrm{u}_9 \\
\end{pmatrix}.
\]
 Then, for each signal in   $\mathds{B}_g$, we construct a constellation that is scaled by the monomial set $\mathcal{G}(\mathds{A}_g)$, i.e, the signals   $\mathbf{X}_{2}^{2,4}$ in $\mathds{B}_2$  uses the monomial   $\mathcal{G}(\mathds{A}_2)$, resulting in the  signal constellation
\begin{equation}
\sum_{v\in \mathcal{G} (\mathds{A}_g)} v\mathbb{Z}_Q.
\end{equation}
Focusing on the users of Example 1, we want to assess whether the
interfering signals have been aligned, and if the requested
subfiles arrive with independent channel coefficients, so that
the decodability is guaranteed. Starting with $\mathrm{u}_1$, the received constellation for the desired signals  $\mathbf{X}_{1}^{1,2}$ $\mathbf{X}_{1}^{1,3}$,  $\mathbf{X}_{1}^{1,4}$, $\mathbf{X}_{2}^{1,2}$, $\mathbf{X}_{2}^{1,3}$ and $\mathbf{X}_{2}^{1,4}$ is given as follow
\begin{equation}
\label{4}
\begin{split}
 C_D&=h_{1,1}\sum_{v\in \mathcal{G} (\mathds{A}_6)} v\mathbb{Z}_Q +h_{1,1}  \sum_{v\in \mathcal{G} (\mathds{A}_4)} v\mathbb{Z}_Q+ h_{1,1} \sum_{v\in \mathcal{G} (\mathds{A}_5)} v\mathbb{Z}_Q \\ &+h_{1,2}  \sum_{v\in \mathcal{G} (\mathds{A}_9)} v\mathbb{Z}_Q +h_{1,2}  \sum_{v\in \mathcal{G} (\mathds{A}_7)} v\mathbb{Z}_Q + h_{1,2} \sum_{v\in \mathcal{G} (\mathds{A}_8)} v\mathbb{Z}_Q.
\end{split}
\end{equation}
The received constellation for the interfering signals  $\mathbf{X}_{1}^{2,3}$ $\mathbf{X}_{2}^{2,3}$,  $\mathbf{X}_{1}^{2,4}$, $\mathbf{X}_{2}^{2,4}$, $\mathbf{X}_{2}^{3,4}$ and $\mathbf{X}_{2}^{3,4}$ is given by
\begin{equation}
\label{5}
\begin{split}
 C_I&=h_{1,1}\sum_{v\in \mathcal{G} (\mathds{A}_1)} v\mathbb{Z}_Q +h_{1,2} \sum_{v\in \mathcal{G} (\mathds{A}_1)} v\mathbb{Z}_Q+ h_{1,1}\sum_{v\in \mathcal{G} (\mathds{A}_2)} v\mathbb{Z}_Q \\ &+h_{1,2}  \sum_{v\in \mathcal{G} (\mathds{A}_2)} v\mathbb{Z}_Q + h_{1,1} \sum_{v\in \mathcal{G} (\mathds{A}_3)} v\mathbb{Z}_Q +h_{1,2}  \sum_{v\in \mathcal{G} (\mathds{A}_3)} v\mathbb{Z}_Q.
\end{split}
\end{equation}
Equation \eqref{5} proves that every two interfering signals, one from each EN, i.e., the first two terms in \eqref{5},  have collapsed into the same constellation space. Also, since the monomials $\mathcal{G} (\mathds{A}_1)$, $\mathcal{G} (\mathds{A}_2)$ and $\mathcal{G} (\mathds{A}_3)$ do not  overlap and linear independence is obtained, the interfering signals will align into $I=3$ different subspaces. 

We can  also see   in \eqref{4} that the  monomials do not align, and  rational  independence is guaranteed (with high probability) and the desired signals will be received over 6 different subspaces. Since the monomials form different constellations, $C_D$ and $C_I$, whose terms are functions of different channel coefficients,
we can assert that these monomials do not overlap. Hence, we can claim that real IA is achieved, and each user can achieve a DoF of $d = \frac{\text{6}}{\text{9}}= \frac{\text{2}}{\text{3}}$. 
In general, the   achievable DoF per user is given by
\begin{equation}
 d = \frac{r {L-1\choose t} }{{L-1\choose t} (r-1)+{L\choose t+1} }.
\end{equation}
Thus, our scheme guarantees that  the desired signals at each user will be received in $r {L-1\choose t}$ different subspaces, and each $r$ interfering  signals will be aligned into the same subspace, i.e., one from each EN, resulting in  a total of $I= {L\choose t+1}- {L-1\choose t}$ interference subspaces.

When $t=L-1$, the number of interference signals at each user is $I=0$. Hence, we just transmit the constellation points corresponding to each signal. We are sure that the decodability is guaranteed since all channel coefficients are i.i.d. according to a continuous  distribution. As a result, each user will be able to achieve a DoF of  $d$=$1$.

User $\mathrm{u}_k$ utilizes  its memory $Z_k$ to extract  the pieces $f^i_{k, \mathcal{T}}$ for  $i\in  \mathcal{N}_k$ and $\mathrm{Index}(i,k)\not \in  \mathcal{T}$. Therefore, user $\mathrm{u}_k$ reconstructs $f^i_{k}$, and
decodes its requested file $W_{k}$. In Example 1, $\mathrm{u}_1$ utilizes its memory $Z_1$ in  TABLE \ref{Table:a} to extract $f^i_{1, \mathcal{T}}$, for $i=1,2$, and $\mathcal{T}=\{2, 3, 4\}$. Hence, $U_1$ reconstructs $f^1_{1}$ and $f^2_{1}$,  and decodes its requested file   $W_{1}$; and similarly for the remaining users. We have  $R_2= 2 {3\choose 1} \frac{1}{2{4\choose 1}}=\frac{3}{4}$.   Thus, the edge NDT  from ENs to the   users   is equal to  $\delta_E (N, M, \rho) = \frac{9}{8}$, while the total NDT  is $\delta (N, M, \rho)=\frac{3}{4\rho}+\frac {9}{8}$. In the general case, 
the  NDT from the  ENs to the users is given by
\begin{equation}
 \delta_E=\frac{{L-1\choose t}(r-1) + {L\choose t+1} }{r{L\choose t} }=\frac{L-t}{r}\left(\frac{r-1}{L} + \frac{1}{t+1} \right) .
\end{equation}


\section{conclusions}
We have studied centralized caching and delivery over a  partially-connected F-RAN with a specified network topology between the ENs and the users. We have proposed a coded caching and delivery scheme that exploits real IA for a receiver connectivity of $r=2$; that is,  when each user can be served by two ENs, or for any receiver connectivity when the user cache capacities are sufficiently large. We have derived the achievable NDT for this scheme, and showed that, increasing receiver connectivity for the same number of ENs and users will   reduce the NDT   for the specific cache capacity values considered, while the amount of reduction depends on the fronthaul rate. The former result follows thanks to the real IA scheme used, which carefully takes care of the interference, and thus, additional connectivity provides better delivery over the edge network, rather than increasing the interference. The latter result is due to the fact  that the size of the transmitted data through each fronthaul link  for the network with higher connectivity  is less than that of the network  with  lower connectivity; and hence, the fronthaul rate helps improve the performance of the latter network more, resulting in a relatively smaller improvement.


\nocite{*}
\bibliographystyle{IEEEtran}
\bibliography{IEEEabrv,mybibfile}
\end{document}